\documentclass[a4paper]{jpconf}
\usepackage{graphicx}

\begin{document}
\title{Centrality dependence of elliptic flow of multi-strange hadrons in Au+Au collisions at $\sqrt{s_{NN}}$ = 200 GeV}

\author{Shusu Shi (for the STAR collaboration)}

\address{Key Laboratory of Quarks and Lepton Physics (MOE) and Institute of Particle Physics, Central China Normal University, Wuhan, 430079, China}

\ead{shishusu@gmail.com}

\begin{abstract}
We present recent results of the mid-rapidity elliptic flow ($v_2$) for multi-strange hadrons and the $\phi$ meson as a function of centrality 
in Au + Au collisions at the center of mass energy $\sqrt{s_{NN}}$ = 200 GeV. The transverse momentum dependence of $\phi$ and $\Omega$ $v_2$ 
is similar to that of pion and proton, indicating that the heavier strange ($s$) quark flows as strongly as the lighter up ($u$) and down ($d$) quarks. 
These observations constitute a clear piece of evidence for the development of partonic collectivity in heavy-ion collisions at the top RHIC energy. 
In addition, the mass ordering of $v_2$ breaks between the $\phi$ and proton at low transverse momenta in the 0-30\% centrality bin, possibly 
due to the effect of late hadronic interactions on the proton $v_2$.
\end{abstract}

\section{Introduction}

In high energy nucleus nucleus collisions, the produced particles are anisotropic in momentum space.
The elliptic flow, $v_{2}$, which is the second Fourier coefficient of the azimuthal distribution of produced particles with respect
to the reaction plane, is defined as $v_{2}=\langle\cos 2(\varphi-\Psi)\rangle$, where $\varphi$ is the azimuthal angle of produced particle and $\Psi$
is the azimuthal angle of the reaction plane. It has been shown that elliptic flow is sensitive to the early stage of heavy-ion collisions and
equation of state of the formed system~\cite{review}. However, early dynamic information might be obscured by later hadronic rescatterings~\cite{hyrdo_cascade}.
Multi-strange hadrons and the $\phi$ meson are believed to be less sensitive to hadronic rescatterings in the late stage of collisions, as their
freeze-out temperatures are close to the phase transition temperature and
their hadronic interaction cross sections are expected to be small~\cite{white, multistrange1, multistrange2}.
Previous measurements of $\phi$ and $\Omega$ $v_{2}$ from STAR were limited by statistics~\cite{old_results1, old_results2}, thus the $p_T$ and centrality dependence of $\phi$ and $\Omega$ $v_{2}$
are not clear. Furthermore, with high precision measurements of $\phi$ meson $v_2$, one can compare it with proton $v_2$ in the low $p_T$ region.
It may provide information on the effect of hadronic rescatterings in the late stage of the collision.

\section{Results and Discussions}

In these proceedings, $v_2$ measurements of multi-strange hadrons from the STAR experiment at $\sqrt{s_{NN}}$ = 200 GeV Au + Au collisions
are presented. About 730 million minimum bias events recorded by STAR in 2010 and 2011 were used in the analysis. The Time Projection Chamber (TPC)
was used for centrality definition and event plane determination. The centrality was determined by the number of tracks from the pseudorapidity region $|\eta|\le 0.5$.
The particle identification (PID) was achieved via energy loss in TPC and flight time in the Time of Flight (TOF). The multi-strange hadrons and the $\phi$ meson 
were reconstructed though the following decay channels:
$\phi$ $\rightarrow$ $\it{K}^{+}$ + $\it{K}^{-}$,
$\Xi^{-}$ $\rightarrow$ $\Lambda$ + $\pi^{-}$ ($\overline{\Xi}^{+}$ $\rightarrow$ $\overline{\Lambda}$ + $\pi^{+}$) and
$\Omega^{-}$ $\rightarrow$ $\Lambda$ + $\it{K}^{-}$ ($\overline{\Omega}^{+}$$\rightarrow$ $\overline{\Lambda}$ + $\it{K}^{+}$).
The $\eta$ sub-event plane method is used for the $v_2$ measurement~\cite{method} . In this method, one defines the event flow vector for each
particle based on particles measured in the opposite hemisphere in pseudorapidity. An $\eta$ gap of $|\Delta\eta| >$ 0.1 between positive and negative
pseudorapidity sub-events is introduced to suppress non-flow effects. 

\begin{figure*}[ht]
\vskip 0cm
\begin{center} \includegraphics[width=0.7\textwidth]{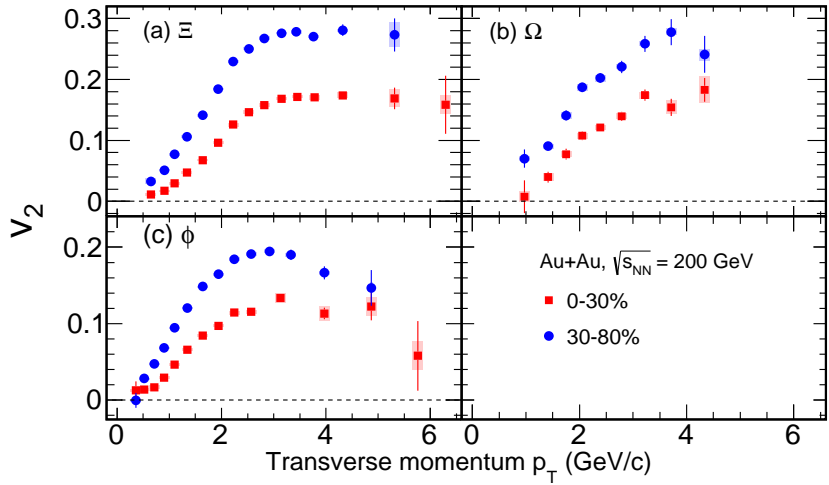}\end{center}
\caption{The $v_{2}$ as a function of $p_{T}$ at
    midrapidity ($|y|<1.0$) for  (a) $\Xi^{-}$ + $\overline{\Xi}^{+}$ (b)
    $\Omega^{-}$ + $\overline{\Omega}^{+}$ and (c) $\phi$ in Au + Au
    collisions at $\sqrt{s_{NN}}$ = 200 GeV for 0-30$\%$ and
    30-80$\%$ centrality. The systematic uncertainties are shown by shaded boxes and the statistical uncertainties by vertical lines. }
\label{figure1}
\end{figure*}

Figure~\ref{figure1}~\cite{paper} shows the elliptic flow  $v_{2}$ as a function of transverse momentum $p_{T}$ at  midrapidity ($|y|<1.0$) for  
(a) $\Xi^{-}$ + $\overline{\Xi}^{+}$, (b) $\Omega^{-}$ + $\overline{\Omega}^{+}$ and (c) $\phi$ in Au + Au collisions at $\sqrt{s_{NN}}$ = 200 GeV for 0-30$\%$ and 30-80$\%$ centrality. 
A clear centrality dependence of $v_{2}(p_{T} )$ is observed for multi-strange hadrons and $\phi$ meson, which is similar to that of light and
strange hadrons previously measured by the RHIC experiments~\cite{star_flow, phenix_flow}. The $v_{2}$ values are
larger in peripheral collisions (30-80$\%$ centrality) than those in central collisions (0-30$\%$ centrality).
This is consistent with an interpretation in which the anisotropy in final momentum space is driven by the anisotropy in initial position space.

\begin{figure*}[ht]
\vskip 0cm
\begin{center} \includegraphics[width=0.8\textwidth]{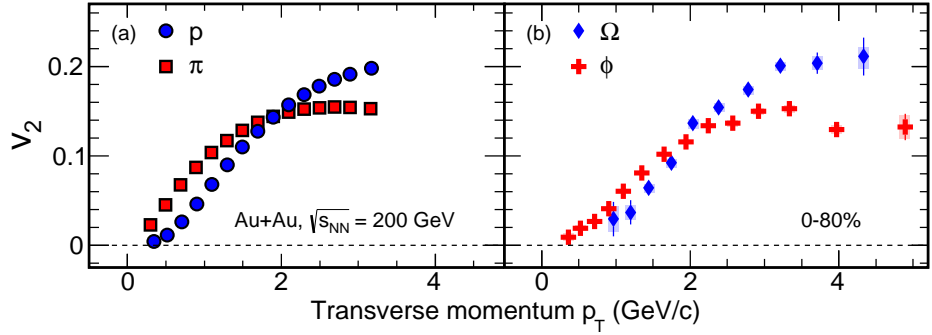}\end{center}
\caption{The $v_{2}$ as function of $p_{T}$ for $\pi$, $p$ (a) and $\phi$, $\Omega$ (b) in Au+Au 
collisions at $\sqrt{s_{NN}}$ = 200 GeV for 0-80$\%$ centrality. The systematic uncertainties are shown by the shaded boxes while vertical lines represent the statistical uncertainties.}
\label{figure2}
\end{figure*}

Figure~\ref{figure2}~\cite{paper} shows the $v_{2}$ as a function of $p_{T}$ for $\pi$, $p$ (panel (a)) and $\phi$, $\Omega$ (panel (b)) in Au + Au collisions at
$\sqrt{s_{NN}}$ = 200 GeV for 0-80$\%$ centrality. 
A comparison between $v_{2}$ of $\pi$ and $p$, consisting of up ($\it u$) and down ($\it d$) light constituent quarks is shown in panel (a). 
Correspondingly, panel (b) shows a comparison of $v_{2}$ of $\phi$ and $\Omega$ containing $\it s$ constituent quarks. 
This is the first measurement of $\Omega$ baryon $v_2$ up to 4.5 GeV/$c$ with high precision.
In the low $p_{T}$ region ( $p_{T}$$<$2.0 GeV/$c$), the $v_{2}$ of  $\phi$ and $\Omega$ follows mass ordering.
At intermediate $p_{T}$  ( 2.0$<$$p_{T}$$<$5.0 GeV/$c$), a baryon-meson separation is observed.
It is evident that the $v_{2}(p_{T}) $ of hadrons consisting only of strange constituent quarks ($\phi$ and $\Omega$)  is similar to that of light hadrons, $\pi$ and $p$. 
However the $\phi$ and $\Omega$ do not participate strongly in the hadronic interactions, because of the smaller hadronic cross sections compared to $\pi$ and $p$. 
It suggests that most of the collectivity is developed during the
partonic phase in Au + Au collisions at $\sqrt{s_{NN}}$ = 200 GeV.

\begin{figure*}[ht]
\vskip 0cm
\begin{center}\includegraphics[width=0.5\textwidth]{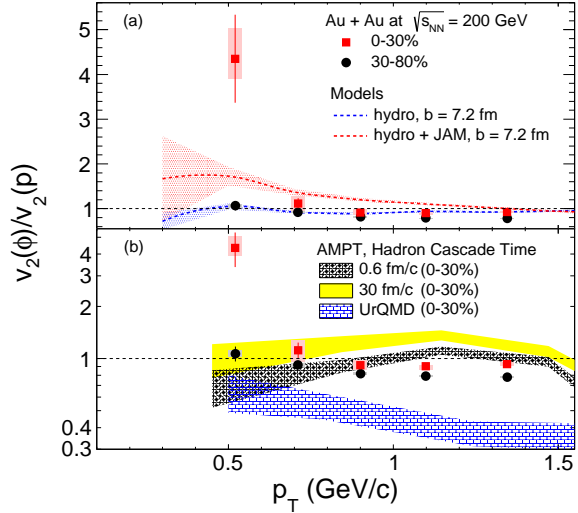}\end{center}
\caption{The ratio of $v_{2}(\phi)$ to $v_{2}(p)$ as function of $p_{T}$ in Au + Au collisions at
$\sqrt{s_{NN}}$ = 200 GeV for  0-30$\%$ and 30-80$\%$ centrality. 
Shaded bands are the  systematic uncertainties and vertical lines are the statistical uncertainties. 
The bands in panel (a) and (b) represent the hydro and transport model calculations for
$v_{2}(\phi)/v_{2}(p)$, respectively. }
\label{figure3}
\end{figure*}

Hydrodynamical model calculations predict that $v_{2}$ as a function of $p_{T}$ for different particle species follows mass ordering, 
where the $v_{2}$ of heavier hadrons is lower than that
of lighter hadrons~\cite{hydro}. The identified hadron $v_{2}$ measured in experiment indeed proves the mass ordering in the
low $p_{T}$ region ($p_{T}$$<$1.5 GeV/$c$)~\cite{star_flow}.
Recently, Hirano {\it et al.} predict the mass ordering of $v_{2}$ could be broken
between $\phi$ mesons and protons at low $p_{ T}$ ($p_{T}$$<$1.5 GeV/$c$)
based on a model with ideal hydrodynamics plus hadron cascade process~\cite{hyrdo_cascade}.
As the model calculations assign a smaller hadronic cross section for $\phi$ mesons compared to protons, 
the broken mass ordering is regarded as the different hadronic rescattering contributions on the $\phi$ meson and
proton $v_2$.
Figure~\ref{figure3}~\cite{paper} shows the ratios of $\phi$ $v_{2}$ to proton $v_{2}$ from model calculations and experimental data. 
This ratio is larger than unity at $p_{T}$ $\sim$ 0.5 GeV/$c$ for 0-30$\%$ centrality. It indicates
breakdown of the expected mass ordering in that momentum range. This could be due to a large  effect of hadronic
rescattering on the proton $v_{2}$, which qualitatively agrees with hydro + hadron cascade calculations indicated by the shaded red band in panel (a) of
Fig.~\ref{figure3}. 
A centrality dependence of $v_{2}(\phi)$ to $v_{2}(p)$ ratio is observed in the experimental data. Namely, 
the breakdown of mass ordering of $v_{2}$ is more pronounced in 0-30$\%$
central collisions than in 30-80$\%$ peripheral collisions. 
In the central events,  both hadronic and partonic interactions are stronger than in peripheral events.
Therefore, 
the larger effect of late stage hadronic interactions relative to the partonic collectivity
produces a greater breakdown of mass ordering in the 0-30$\%$ centrality data than in the 30-80$\%$.
This observation indirectly supports the idea that the $\phi$ meson has a smaller hadronic interaction cross section.  
The ratio of $\phi$ $v_{2}$ to proton $v_{2}$ was also studied by using the transport models AMPT~\cite{ampt} and UrQMD~\cite{urqmd}. 
The panel (b) of Fig.~\ref{figure3} shows the $v_{2}(\phi)$ to $v_{2}(p)$ ratio for 0-30$\%$ centrality from AMPT and UrQMD models.
The black shaded band is from AMPT with a hadronic cascade time of 0.6 fm/$c$ while the yellow
band is for a hadronic cascade time of 30 fm/$c$.  Larger hadronic cascade time is equivalent to stronger hadronic interactions.
It is clear that the $v_{2}(\phi)/v_{2}(p)$ ratio increases with increasing hadronic cascade time. 
This is attributed to a decrease in the proton $v_{2}$ due to an  increase in hadronic re-scattering
while the $\phi$ meson $v_{2}$ is less affected. The ratios from the UrQMD model
are much smaller than unity (shown as a blue shaded band in the panel (b) of Fig.~\ref{figure3}).
The UrQMD model lacks partonic collectivity, thus the $\phi$ meson $v_{2}$ is not fully developed.

\section{Summary}
In summary, high-statistics  elliptic flow measurements for multi-strange hadrons
($\Xi$ and $\Omega$) and  $\phi$ meson as a function of centrality and $p_T$ were reported in Au + Au
collisions at $\sqrt{s_{NN}}$ = 200 GeV. The $p_{T}$ dependence of 
$\phi$ and $\Omega$ $v_{2}$ is observed to be similar to that of light hadrons, $\pi$ and $p$,
indicating that most of the collectivity is developed in the initial
partonic phase for light and strange hadrons. 
The comparison between the $\phi$ and $p$ $v_{2}$ at low
$p_{T}$ shows that there is a possible violation of hydro-inspired mass ordering between $\phi$ and
$p$. Model calculations suggest that the
$p_{T}$ dependence of $v_{2}(\phi)$ to $v_{2}(p)$ ratios can be qualitatively explained by
the different effects of late-stage hadronic interactions on the
$\phi$ and proton $v_{2}$.

\section{Acknowledgments}
This work was supported in part by the National Natural Science Foundation of China under grant No. 11475070, 
National Basic Research Program of China (973 program) under grand No. 2015CB8569 and 
self-determined research funds of CCNU from the colleges' basic research and operation of MOE under grand No. CCNU15A02039.


\begin{thebibliography}{00}
\bibitem{review} S. A. Voloshin, A. M. Poskanzer and R. Snellings, in Landolt-Boernstein, Relativistic Heavy Ion Physics, Vol. 1/23, p. 5-54 (Springer-Verlag, 2010). 
arXiv:0809.2949.
\bibitem{hyrdo_cascade} T. Hirano  {\it et al.},  Phys. Rev. \textbf{C 77}, 044909 (2008).
\bibitem{white} J. Adams {\it et al.}, Nucl. Phys. \textbf{A 757} 102 (2005).
\bibitem{multistrange1} A. Shor, Phys. Rev. Lett. \textbf{54}, 1122 (1985).
\bibitem{multistrange2} H. van Hecke, H. Sorge and N. Xu, Phys. Rev. Lett. \textbf{81}, 5764 (1998).
\bibitem{old_results1} J. Adams {\it et al.}, Phys. Rev. Lett. \textbf{95}, 122301 (2005).
\bibitem{old_results2} S. S. Shi (for the STAR collaboration), Nucl. Phys. A {\bf 830}, 187c (2009);
Nucl. Phys. A {\bf 862-863}, 263 (2011).
\bibitem{method} A. M. Poskanzer and S. A. Voloshin,  Phys. Rev. \textbf{C 58}, 1671 (1998).
\bibitem{paper} L. Adamczyk {\it et al.}, arXiv:1507.05247 (2015).
\bibitem{star_flow} B. I. Abelev {\it et al.}, Phys. Rev. \textbf{C 77}, 054901 (2008).
\bibitem{phenix_flow} A. Adare {\it et al.}, Phys. Rev. \textbf{C 85}, 064914 (2012).
\bibitem{hydro} P. Huovinen  {\it et al.}, Phys. Lett. \textbf{B 503}, 58 (2001). 
\bibitem{ampt} Z.-W. Lin {\it et al.},   Phys. Rev. \textbf{C 72}, 064901 (2005).
\bibitem{urqmd} S. A. Bass {\it et al.},  Prog. Part. Nucl. Phys. \textbf{41}, 255 (1998).

\end{thebibliography}
\end{document}